# Do they agree? Bibliometric evaluation vs informed peer review in the Italian research assessment exercise

Alberto Baccini

(Dept. of Economics and Statistics, University of Siena. Italy
Piazza San Francesco 7, 53100 Siena, alberto.baccini@unisi.it; tel. +390577233076)

Giuseppe De Nicolao

(Dept. of Electrical, Computer and Biomedical Engineering, University of Pavia, Italy)

**ABSTRACT**

During the Italian research assessment exercise, the national agency ANVUR performed an experiment to assess agreement between grades attributed to journal articles by informed peer review (IR) and by bibliometrics. A sample of articles was evaluated by using both methods and agreement was analyzed by weighted Cohen's kappas. ANVUR presented results as indicating an overall "good" or "more than adequate" agreement. This paper re-examines the experiment results according to the available statistical guidelines for interpreting kappa values, by showing that the degree of agreement (always in the range 0.09-0.42) has to be interpreted, for all research fields, as unacceptable, poor or, in a few cases, as, at most, fair. The only notable exception, confirmed also by a statistical meta-analysis, was a moderate agreement for economics and statistics (Area 13) and its sub-fields. We show that the experiment protocol adopted in Area 13 was substantially modified with respect to all the other research fields, to the point that results for economics and statistics have to be considered as fatally flawed. The evidence of a poor agreement supports the conclusion that IR and bibliometrics do not produce similar results, and that the adoption of both methods in the Italian research assessment possibly introduced systematic and unknown biases in its final results. The conclusion reached by ANVUR must be reversed: the available evidence does not justify at all the joint use of IR and bibliometrics within the same research assessment exercise.

## 1. Introduction

One of the most discussed issue in research evaluation and policy literature is about the agreement between peer review and bibliometric indicators. When these generic expressions are clearly defined, a plurality of research issues emerges such as the question of the agreement between the judgements of expert peer-reviewers commenting articles at the time of their publication and some form of article-level citation metrics (Allen et al. 2009; Berghmans et al. 2003); or the analysis of the correlation between peer ratings assigned to individual researchers or research groups, and bibliometric indicators calculated for these same researchers or groups (van Raan 2006; Lovegrove and Johnson 2008; Koenig 1983; Rinia et al. 1998; Aksnes and Taxt 2004). Because of its research policy relevance, the most discussed issue is the comparison of results of national research assessment exercises obtained by using peer review, and results obtained by using bibliometric indicators. This stream of literature was carefully reviewed by (Wouters et al. 2015) in the context of the "Independent review of the role of metrics in research assessment and management", commissioned by the Higher Education Funding Council for England. In the context of this review an analysis of the Research Excellence Framework (REF) 2014 results and metric indicators was performed, by providing probably the most important piece of evidence available for this topic (HEFCE 2015). Scores assigned by peer reviewers to articles submitted to REF are compared with 15 bibliometric and altmetric indicators of research performance defined at a paper level, such as citation count, field-weighted citation impact, percentile of highly cited publications, SCImago Journal Rank of the journal in which an article is published and so on. "This work –according to the authors- has shown that individual metrics give significantly different outcomes from the REF peer review process, showing that metrics cannot provide a like-for-like replacement for REF peer review" (HEFCE 2015). Similar conclusions were reached by using REF data and bibliometric indicators defined at a departmental level (Mryglod et al. 2015).

This overwhelming evidence is challenged by results obtained in the Italian research assessment exercise, the so called Valutazione della Qualità della Ricerca (Evaluation of Research Quality) 2004-2010 (hereinafter indicated by its Italian acronym VQR) managed by the Italian Agency for the evaluation of the university and research (ANVUR) and concluded in 2013. Italian VQR adopted a "dual system of evaluation" (Ancaiani et al. 2015) in which both peer review and bibliometrics were considered. "In order to validate the use of the dual system" (Ancaiani et al. 2015), ANVUR performed a massive comparison among peer review and bibliometric indicators as a part of the VQR (we will refer hereafter to this comparison as "the experiment"). Indeed, more than nine thousands journal articles (the 10% of total) submitted for the assessment were evaluated by ANVUR by using both bibliometrics and peer review. ANVUR summarized results of the experiment by stating that "In the complex ... there is *a more than adequate concordance* between evaluation carried out through peer review and through bibliometrics. This result fully justifies the choice made at VQR [...] to use both techniques of assessment" (ANVUR 2013 Appendix B, p. 25-26, italics added).

The aim of this paper is to challenge data and conclusions reached by ANVUR (ANVUR 2013). Since raw data were not made available by ANVUR,[1] it was impossible for us to control in details the experiment and to reproduce the results published and by ANVUR and by its collaborators in different outlets. We had to perform our analysis by using only data already published in the *Final Report* of the VQR and in the *Final Reports* of the research areas as described below. In particular, we re-analyzed data by using available statistical guidelines and by performing a statistical meta-analysis of the experiment. The rest of the paper is organized as follows. Section 2 contains a description of Italian VQR functional to the understanding of the rest of the paper. Section 3 describes in some details ANVUR's experiment and re-analyzes data on the basis of available statistical guidelines for the techniques used by ANVUR. Section 4 contains a statistical meta-analysis of the experiment, by showing that only in a research field (Economics and statistics) there was evidence of a good agreement between results reached through peer review and bibliometric indicators. Section 5 shows that the good agreement in this research field can be explained by an experimental protocol which differs substantially from the one adopted in all the other research areas. After having concluded that bibliometrics and peer review produced different evaluations, the general interest of these results is discussed in the conclusions.

---

[1] Data had been requested to the President of the ANVUR with a mail sent the 10th February 2014. We have not received yet a reply.

**2. The Italian research assessment exercise**

It is impossible to analyze the experiment without considering the general design of the Italian research assessment exercise (for a more detailed discussion see Abramo and D'Angelo 2015; Baccini 2016). VQR was the second massive research assessment exercise for universities and research institutions realized in Italy. Since the first exercise was realized with different rules and methodologies, results of this second exercise cannot be compared with the previous ones. It covered the time period 2004-2010. The final report of the exercise was published in July 2013 (ANVUR 2013); it consists of a general report with 7 appendices, hereafter indicated as *Final Report*, and 14 *Area reports* with their appendices.[2] Indeed VQR was organized in 14 widely defined research Areas, labeled with numbers from 1 to 14.[3] These area originates from the traditional classification of research areas adopted in Italy for the election of the representatives in the Italian National University Council. For each area, an evaluation panel, the so called "GEV", was established with a number of panelists proportional to the number of research outputs to be evaluated; the total number of panelists was 450. Each panel was organized in sub-panels, called "sub-GEV", specialized for specific research fields, so a total of 44 sub-panels were defined. Sub-panels directly managed and evaluated subsets of research products submitted for evaluation in their area of expertise.

The minister stated in a ministerial decree[4] the general structure of the VQR and also the merit classes which had to be adopted by panels. ANVUR defined the general structure of the evaluation, delegating further decisions about evaluation methods and technicalities to the area panels. This complex arrangement resulted in a division between the so called "bibliometric areas", (Areas from 1 to 9, that is *hard* sciences, engineering and life sciences) where evaluation was conducted, mainly but not exclusively, through bibliometrics, and the so called "non-bibliometric areas" (Areas 10-12 and 14, that is social science and humanities, excluding economics and statistics) where evaluation was conducted exclusively through peer review. Area 13 (economics and statistics) was an exception: it was also classified as "non bibliometric" and the evaluation was performed through bibliometric instruments but different from the ones used in the other bibliometric areas, as we will describe below.

The aim of the VQR was to evaluate research institutions such as universities or departments, and research areas and fields both at national or institutional levels. Each university, department and research field was synthetically evaluated by calculating the average score obtained by the research outputs submitted by researchers.[5] To this end, all the researchers with a permanent position in an university had to submit three research outputs for evaluation; all researchers enrolled in institution different from universities had to submit six research outputs. Each researcher was classified according to the specific research field in which it is officially enrolled; indeed in Italy each university researcher is classified in one of the 370 research fields into which the 14 Areas listed above are subdivided.[6] Each research work submitted was classified according to the research field of its author. Each research work was then evaluated and received a score. The set of admissible scores, specified by the already cited ministerial decree, derived from a questionable adaptation of the rules of the British Research Assessment Exercise (RAE 2005). Indeed, in the ministerial decree it is stated that for each research field "a scale of values shared by the international scientific community" exists for judging the quality of the research.[7] In particular, four merit classes were therefore defined as:

- Excellent (score 1): the research work is placed in the top 20% of the "scale of values shared by the international community";
- Good (score 0.8), the research work is placed in the 60-80% interval of the scale;
- Acceptable (score 0.5): the research work is in the 50-60% interval of the scale;
- Limited (score 0): the research work is in the lowest 50% of the scale.[8]

---

[2] The *Final Report* and all the *Area Reports* are in Italian. Quotations from these documents are translated by the authors. Appendix A of the Area 13 Report is in English.
[3] The 14 areas are: Mathematics and informatics (Area 1); Physics (Area 2); Chemistry (Area 3); Earth Sciences (Area 4); Biology (Area 5); Medicine (Area 6); Agricultural and Veterinary Sciences (Area 7); Civil Engineering and Architecture (Area 8); Industrial and Information Engineering (Area 9); Antiquities, Philology, Literary studies, Art History (Area 10); History, Philosophy, Pedagogy and Psychology (Area 11); Law (Area 12); Economics and Statistics (Area 13); Political and Social sciences (Area 14).
[4] Minister of Education, decree n. 17, 2011/07/15.

The thresholds used for this classification defined the so-called "VQR distribution rule (20-20-10-50)". The ministerial decree stated also that research work could be also penalized if it did not belong to the typologies admissible for evaluation (score -1), or in case of fraud or plagiarism (-2). Researchers who failed to provide the requested number of research outputs determined a penalization for their institution (a score of -0.5 for each missing research output).

In the non-bibliometric areas, each research output was evaluated by a couple of anonymous reviewers, chosen by one or two members of the sub-area panels. The two referees summarized their judgments on a predefined format, slightly differentiated among the GEVs, that required to indicate separate scores for three dimensions: originality, relevance, internationalization. For each referee, a total score, using rules slightly differentiated among GEVs, was then calculated. The two referees judgments were then summarized and classified in one of the four classes by the GEV members who had chosen the referees, and the final score for the research output finally decided.

In the bibliometric areas (1 to 9), a mix of peer review and bibliometrics was adopted. All research outputs different from articles in journals were evaluated by peer review as in the non-bibliometric areas. Journal articles were instead evaluated through bibliometrics or peer review. The procedure for evaluating journal articles was rather involved. Each article was classified by considering two bibliometric indicators: (i) the total number of citations received in the time window starting from its publication and 31th December 2011, and (ii) the impact factor of the journal in which it was published (ad-hoc bibliometric indicators were also devised by some of the GEVs). The main sources for citations and journal impact indicators were respectively Web Of Science and Thomson Reuters *Journal Citation Reports* (2010). The world distribution for subject categories and years of these two indicators was considered. Normalizations were not introduced neither for citation windows nor for research fields. For each year and each subject category, the distributions of citations and impact factors were segmented each in four classes according to the percentiles of the VQR distribution rule.[9] So, for citations in a given research field and year, the world distribution of articles was segmented in four classes: the first contained the top 20% of articles for number of citations; the second class contains the articles in the 60-80% range of citations; the third the articles in the range 50-60%; and the fourth class the articles with a number of citation below the median value. Analogously, for each research field considered, journals were segmented according to their impact factors by using the VQR distribution rule. Four by four matrices with the 16 possible combinations of classes of citations and journal impact factors were then defined. Each GEV defined its own algorithms for translating these possible combinations in the final score of the articles. At least 48 different matrices were used for article classification (Baccini 2016). Figure 1 shows one of the matrices adopted by the GEV 5.

---

[5] The resulting university and department rankings were disseminated through a booklet (AA.VV. 2013).
[6] The complete list of the Italian scientific classification is available with the official English translation: https://www.cun.it/uploads/storico/settori_scientifico_disciplinari_english.pdf
[7] Minister of Education, decree n. 17, 2011/07/15, art. 8 comma 4.
[8] This scale of value appears to be operationally meaningless (Baccini 2016); similar critiques are moved also to the British Research Assessment (McNay 2011).

**2004-2008**

**Indicatore bibliometrico**

|   | 1 | 2 | 3 | 4 |
|---|---|---|---|---|
| 1 | A | A | A | IR |
| 2 | B | B | B | IR |
| 3 | IR | C | C | C |
| 4 | IR | D | D | D |

n. di citazioni

**Figure 1.** A bibliometric evaluation matrix adopted by GEV 5. On the horizontal axis the impact factor value is categorized in four classes according to the VQR distribution rule and the same is done for citations on the vertical axis. The letters A, B, C, D denote the four merit classes, while IR means that the research work has to go through Informed Review. Source: Reproduced from (ANVUR 2013)

Each submitted article belonged to one of the 16 squares of the matrix, according to the number of citations received and the impact factor of the journal in which it had been published. Articles were automatically assigned a score, when information about citation and impact factor agreed. For example, if citations were in the top 20% of the articles for its fields, and the journal in the top 20% of journals according to impact factor, the research product was evaluated as excellent by awarding a score of 1 point. When impact factor and citations gave divergent information -high citations and low impact factor or *viceversa*- the corresponding entry in the classification matrix was denoted by "IR" indicating that the article had to be submitted to a process called "informed peer review" (IR). For example, an article published in a journal with an impact factor higher than median, but a very low number of citations, had to be submitted to IR. In these cases, anonymous reviewers were provided with the complete metadata of the article, the number of citations received by the article and the journal impact factor. Reviewers were asked to give a score to the article by considering all the information available.

---

[9] Area 9 used quartiles instead of the percentiles of the VQR distribution rule for defining the journal segments.

Area 13 (Economic Science and Statistics) adopted a different system of bibliometric evaluation, by scoring each article according to the ranking of the journal in which it had been published. Journal rankings were developed directly by the GEV. Journal were classified according to the VQR distribution rule applied to real or imputed indicators for journals (five year impact factor IF5 and article influence score AIS for the year 2010).[10] In this area, journal articles were not subjected to IR.

ANVUR coined the expression "evaluative mix" to denote this complex machinery. It gave rise not only to the adoption of evaluation rules specifically defined for research fields and sub-fields, but also to a peculiar mix of scores obtained for bibliometric areas in part by bibliometrics and in part by informed peer review; and by peer review in non-bibliometric areas. The adoption of this evaluative mix created many problems when evaluation results were used for comparing institutions and research fields. The first major problem was originated in bibliometric areas: the joint distribution of article citations and journal impact factors varies among research fields. As a consequence the VQR distribution rule was not respected for all fields, and because of this in some fields to be classified as excellent or good was simpler that in other fields. ANVUR was aware of this problem but did not address it.[11] Although an explicit warning was not issued at the time of publication of results, there was lack of comparability not only between Areas but also between research fields inside the same research Area.[12] In fact, in 2014 ANVUR published new sets of scores that had been normalized so as to neutralize biases across research fields within the same research Area. Such normalized scores were used, together with the original ones, for the accreditation procedure of PhD courses.[13]

The second major problem was about the bias induced by the adoption of different evaluation techniques. Indeed, if IR produced scores systematically different from the ones produced by bibliometrics, this might have introduced systematic bias in the scoring system. As a consequence, the average scores of an institution might be distorted by the different percentage of scores attributed by IR and by bibliometrics. To address this problem, ANVUR carried out an experiment for assessing the agreement of the two evaluation methods: for Areas 1-9 and 13, a random sample of articles was submitted to both bibliometric and IR evaluation. This experiment is crucial for the overall soundness of the results of the national assessment. Only a good agreement between bibliometric evaluation and evaluation performed by IR may justify the adoption of the two different evaluation methods and preserve the comparability of results among areas, institutions, departments and research fields

**3. Comparing IR and bibliometrics**

This experiment was summarized in Appendix B of the *Final Report* and in Appendix A of each of the ten *Area reports* (1-9,13). Its relevance is testified by the dissemination effort produced by ANVUR collaborators. Indeed the overall results were also published as a working paper in Italian (Cicero et al. 2013) and as a section of a paper (Ancaiani et al. 2015); the Appendix A of the *Area 13 Report,* that is the report of the experiment for economics and statistics, was published also as a working paper which appeared in a nearly identical form in five different working paper series (Bertocchi et al. 2013e, 2013a, 2013b, 2013c, 2013d), and it is now also a "research article" published in a scholarly journal (Bertocchi et al. 2015).

---

[10] The procedure is described in the *Area 13 report* and reproduced in (Bertocchi et al. 2015).
[11] ANVUR published estimates of the bias in the Appendix A of the *Final Report.*
[12] At the same time ANVUR diffused a press release comparing the average score of the 14 areas (http://www.anvur.org/rapporto/files/stampa/TABELLA%201.pdf), that was extensively used by newspapers in their coverage of the news about VQR.
[13] http://www.anvur.org/attachments/article/455/valutazione%20corsi%20dottorato_finale_clean.pdf

As anticipated, about a ten percent of articles were evaluated by both IR and bibliometrics in order to verify the agreement between the two evaluation systems. The basic structure of the experiment as reported by ANVUR is the following.[14] A random sample of 9,199 articles was extracted from the 99,005 submitted for evaluation. The sample was stratified by GEVs and sub-areas.[15] The overall sample represented the 9.3% of the total number of submitted articles. A bibliometric evaluation was assigned to each article according to the criteria defined by the GEVs.[16] Each article was evaluated also through IR. In this case two referees, chosen by two different members of the GEV, were asked to evaluate according "to their subjective perception of the quality of the product in reference to the world distribution of the research products of the scientific field" (ANVUR 2013). Referees were provided with metadata of research articles and with bibliometric data. The two referees summarized their judgments on the predefined format adopted by each sub-GEV by indicating separate scores: originality, relevance, internationalization. For each referee a total score was then calculated. The two scores were finally summarized in a final evaluation "based on algorithms specific for each Area" (Cicero et al. 2013). These algorithms, only partially disclosed, transformed the final evaluation in one of the four merit classes and scores described above.

The bulk of ANVUR experiment consisted in the analysis of the agreement between the evaluation obtained through IR and bibliometric data. The statistical technique adopted by ANVUR was Cohen's kappa (Cohen 1960), the most popular index of interrater agreement for nominal categories (Sheskin 2003). Cohen's kappa is a chance corrected agreement rate between two raters who independently assign a sample of units to two or more categories (nominal scale); it is defined as:

$$\kappa = \frac{p_o - p_e}{1 - p_e}$$

where $p_o$ is the observed proportion of agreement between the raters and $p_e$ is the proportion of agreement expected by chance. The upper limit is $k$=1 occurring when the two raters perfectly agrees; value of $\kappa \leq 0$ indicates that observed agreement is less than or equal to agreement expected by chance (Fleiss et al. 2003). When more than two nominal categories are considered, and when these categories are hierarchically ordered, weighted kappa can be introduced for describing specific patterns of disagreement (Cohen 1968). The weights used in the calculation of Cohen's kappa indicate the seriousness of the disagreement, by giving minimal weight to minimal disagreement and maximum weight to maximum possible disagreement. The most diffused weights are linear weights and quadratic weights (Fleiss et al. 2003). With a large enough number of observations, the sampling distribution of kappa and weighted kappa is approximately normal, enabling the calculation of confidence intervals and a significance test (Sheskin 2003). It is worthwhile to note that the significance test "is generally of little practical value, since a relatively low value of kappa can yield a significant result. In other words, a value such as $k = 0.41$ (in spite of the fact that is statistically significant) may be deemed by a researcher to be too low a level of reliability (i.e. degree of agreement) to be utilized within a practical context" (Sheskin 2003).

---

[14] The exposition is based on Appendice B, par. B.1 of the *Final Report*.
[15] In the Areas 1, 3, 5, 6, 7, 8 the sub-areas corresponded to the sub-GEVs in which the panel organized the evaluation. In Area 2 (physics), the seven sub-areas were defined directly by the GEV by modifying the Web of Science classification; Area 4 was partitioned in four sub-areas corresponding to an administrative classification called "settori concorsuali", that is a classification adopted by Italian government for the recruitment of professors; Area 9 was partitioned in 9 macro-fields defined directly by the GEV; finally, Area 13 adopted a fourfold classification developed directly by the GEV. These pieces of information are drawn from Area reports, but they are not properly disclosed neither in the *Final Report* neither in (Cicero et al. 2013)
[16] It is worthwhile to note that many articles were dropped by the experiment because the bibliometric evaluation resulted in an inconclusive IR value. This induced a distortion in the sample that was not considered in the analysis by ANVUR.

Indeed the main problem is "how to maintain a consistent nomenclature when describing the relative strength of agreement associated with kappa statistics" (Landis and Koch 1977), that is how to judge and describe different ranges of kappa values with respect to the degree of agreement they suggest. Although the meaning of value 0 or 1 is clear, the interpretation of intermediate values is less evident. Different guidelines are available (Fleiss et al. 2003; Landis and Koch 1977; Altman 1991; George and Mallery 2003; Stemler and Tsai 2008) as summarized in Table 1.

**Table 1. Available guidelines for interpreting kappa values**

| **Landis and Kock (1977)** | | **Altman (1991)** | | **George and Mallery (2003)** | |
|---|---|---|---|---|---|
| *K values* | *Description* | *K values* | *Description* | *K values* | *Description* |
| <0.00 | Poor | <0.20 | Poor | <0.51 | Unacceptable |
| 0.00-0.20 | Slight | 0.21-0.40 | Fair | 0.51-0.60 | Poor |
| 0.21-0.40 | Fair | 0.41-0.60 | Moderate | 0.61-0.70 | Questionable |
| 0.41-0.60 | Moderate | 0.61-0.80 | Good | 0.71-0.80 | Acceptable |
| 0.61-0.80 | Substantial | 0.81-1.00 | Very good | 0.81-0.90 | Good |
| 0.81-1.00 | Almost perfect | | | 0.91-1.00 | Excellent |

| **Stemler and Tsai (2008)** | | **Fleiss (2003)** | |
|---|---|---|---|
| *K values* | *Description* | *K values* | *Description* |
| <0.50 | Unacceptable | <0.40 | Poor |
| >0.50 | Acceptable | 0.40-0.75 | Fair to good |
| | | >0.75 | Excellent |

In the case at hand, IR and bibliometric evaluation were considered as two raters expressing different judgement on same articles. These judgements assigned each article to one of the four merit classes. Since the four merit classes are hierarchical ordered, ANVUR decided to implement weighted kappa measures. ANVUR used two sets of weights. The first one is a set of linear weights (0; 0.33; 0.67; 1) where the lowest possible disagreement, that is one merit class disagreements (disagreements between A and B; B and C; C and D) are weighted one third of the strongest possible disagreement, that is the three merit classes disagreement (between A and D). The second set of weights, called VQR-weights, is the one which according to ANVUR should be preferred because it adopts "appropriate weights … associated with the qualitative evaluation" (Bertocchi et al. 2015), as the VQR-weights reproduced the VQR-distribution rule (0; 0.5; 0.8; 1). This last set of weights attributes a decreasing marginal weight to stronger disagreement. A system of weight of this kind appears to be counter-intuitive and probably for this reasons it was never adopted previously in the long-lasting and consolidated stream of literature using Cohen's kappa.

Table 2 presents Cohen's kappas calculated by ANVUR for the 10 Areas and 43 sub-areas.[17] The kappas in the first column are the ones calculated by adopting linear weights; the ones in the second columns are calculated by adopting the VQR-weights. Kappas calculated for the whole sample of articles are respectively 0.32 with linear weights and 0.38 with VQR-weights.[18] That is in the whole sample just a little less or a little more than one third of evaluations reached through bibliometrics and IR are agreements. When linear weights are considered, kappa values for all the areas are in the range 0.1-0.35, with the only exception of a value of 0.54 for Area 13 (Economics and statistics). When VQR-weights are considered, kappas for all the areas are again in the range 0.15-0.35 with the exception of 0.54 for Area 13 (Economics and statistics).[19] When sub-areas and linear weights are considered the minimum kappa is calculated for the nuclear physics (0.095) and the maximum one for economics (0.56); only four sub-areas out of 43 have kappa values greater than 0.40. If sub-areas and VQR-weights are considered, the minimum kappa is calculated for electronic engineering (0.09) and the maximum one for economics (0.56); again only four sub-areas have kappa values greater than 0.40.

---

[17] ANVUR stressed the importance of significance test for Cohen's kappa, by improperly interpreting statistical significance as agreement. In the section of an article reproducing result of the experiment, the statistical significance

Surprisingly enough, ANVUR's *Final Report* described generally these data as indicating "a *good* degree of agreement for the whole sample and for each GEV" [italics added]; in the conclusion of the report, results are summarized by writing that there is "a more than adequate [in Italian "più che adeguata"] agreement between evaluations done by adopting the peer review method and the bibliometric one" (Cicero et al. 2013; ANVUR 2013). This phrase is repeated *verbatim* in the conclusions of all the *Area Reports*, irrespective of the range of values commented.[20] Results of the experiment are also presented and commented as "giving evidence of a significant degree of concordance among peer review and bibliometric evaluations" (Ancaiani et al. 2015). In reference to the Area 13 results, in the Area report the degree of agreement is also considered as "more than adequate" (ANVUR 2013). This phrase is slightly reworded in the working paper versions: "there is remarkable agreement between bibliometric and peer review evaluation" (Bertocchi et al. 2013d, 2013e, 2013a, 2013b, 2013c). And the same data are also interpreted in a policy oriented paper as revealing that "informed peer review and bibliometric analysis produce similar evaluations" (Bertocchi et al. 2014). They are finally commented as "fair to good agreement" in (Bertocchi et al. 2015).[21]

The use made by ANVUR of expressions such as "good degree of agreement" and "more than adequate" appears to be misleading when existing guidelines reported in Table 1 are considered. According to these guidelines, results of the experiment, that is almost all the values for kappas listed in Table 2, are indicative of agreement that can be described as "unacceptable", or alternatively as "poor" or "fair", for nearly all the Areas and sub-areas. The only exceptions are Area 13 (Economics and statistics), and four sub-areas, three of which are defined inside Area 13, with an agreement that can be described as "acceptable" or alternatively as "fair to good" or "moderate". Overall results should be interpreted as indicating that the hypothesis of a

---

of kappa values was even exchanged for agreement: "kappa is always statistically different from zero, showing that there is a fundamental agreement" (Ancaiani et al. 2015).

[18] For the whole sample and VQR-weights, a value of 0.3441 is reported in (Ancaiani et al. 2015).

[19] For Area 13 and VQR-weights, a value of 0.6104 is reported in the Appendix B of the ANVUR Final Report and also in (Cicero et al. 2013) and (Ancaiani et al. 2015). The value of 0.61 appears inconsistent when the other kappas calculated for the sub-areas of Area 13, reported in Table 2, are considered. The value of 0.54 that we used in this paper is drawn directly from the Area 13 report and reproduced also in (Bertocchi et al. 2015).

[20] In the conclusion of the Area 9 report, that phrase is followed by the contradictory statement that: "The degree of concordance between peer evaluations and bibliometric evaluations is moderate (in Italian: "moderato") in near all the sub-areas, while it results rather high (in Italian: "piuttosto elevato") for informatic engineering" (ANVUR 2013).

[21] (Bertocchi et al. 2015) introduced references to the relevant literature which were not presented in the working paper versions of the paper. They wrote "Since the most common scales to subjectively assess the value of kappa mention "adequate" and "fair to good", these are the terms we use in the paper." Really, the term "adequate" is not used in the relevant literature, but it is adopted by the ANVUR only in its reports.

moderate (or stronger than moderate) agreement between IR and bibliometrics is not fulfilled, with the exception of the evidence found for Area 13 which appears to be favorable to the hypothesis of a moderate agreement. The question at this point is to verify if the results for Area 13 are really different from the results of the other areas and sub-areas. To this end, we performed a meta-analysis on the VQR results.

**Table 2. Weighted kappas values for Areas and sub-areas.**

| | Sample | Linear weighted kappas | VQR weighted kappas |
|---|---|---|---|
| **Area 1 Mathematics and Informatics** | **631** | **0,3176** | **0,3173** |
| Informatics | 164 | 0,3794 | 0,3896 |
| Mathematics | 121 | 0,3218 | 0,3102 |
| Analysis and probability | 179 | 0,2551 | 0,2755 |
| Applied Mathematics | 167 | 0,2426 | 0,2403 |
| **Area 2 Physics** | **1412** | **0,2302** | **0,2515** |
| Experimental Physics | 139 | 0,1957 | 0,2049 |
| Theoretical Physics | 499 | 0,2428 | 0,2559 |
| Physics of matter | 349 | 0,1862 | 0,2099 |
| Nuclear and sub-nuclear physics | 45 | 0,0951 | 0,1001 |
| Astronomy and astropyisics | 270 | 0,2708 | 0,3048 |
| Geophysics | 28 | 0,3671 | 0,3975 |
| Applied physics, teaching and history | 82 | 0,2153 | 0,2715 |
| **Area 3 Chemistry** | **927** | **0,2246** | **0,2296** |
| Analitical chemistry | 276 | 0,2261 | 0,2192 |
| Inorganic and industrial chemistry | 283 | 0,2024 | 0,2158 |
| Organic and farmaceutical chemistry | 368 | 0,2304 | 0,2368 |
| **Area 4 Earth sciences** | **458** | **0,2776** | **0,2985** |
| Geochemistry etc. | 123 | 0,287 | 0,2996 |
| Structural geology | 96 | 0,1891 | 0,1932 |
| Applied geology | 56 | 0,2736 | 0,3375 |
| Geophysics | 183 | 0,277 | 0,3125 |
| **Area 5 Biology** | **1310** | **0,3287** | **0,3453** |
| Integrated biology | 325 | 0,3451 | 0,3648 |
| Morfo-functional sciences | 216 | 0,3629 | 0,3775 |
| Biochemistry and molecular biology | 410 | 0,2998 | 0,304 |
| Genetics and pharmacology | 359 | 0,296 | 0,3248 |
| **Area 6 Medicine** | **1984** | **0,303** | **0,3351** |
| Experimental medicine | 347 | 0,2407 | 0,2602 |
| Clinical medicine | 968 | 0,2883 | 0,3128 |
| Surgical sciences | 554 | 0,3368 | 0,385 |
| Public health | 115 | 0,2023 | 0,2176 |
| **Area 7 Agricoltural and Veterinary sciences** | **532** | **0,2776** | **0,3437** |
| Agricoltural sciences | 387 | 0,2741 | 0,3354 |
| Veterinary | 145 | 0,2747 | 0,3514 |
| **Area 8 Civil engineering and Architecture** | **225** | **0,1994** | **0,2261** |
| Infrastructural engineering | 99 | 0,2106 | 0,2052 |
| Structural engineering | 126 | 0,2037 | 0,2544 |
| **Area 9 Industrial and Information Engineering** | **1130** | **0,1615** | **0,171** |
| Mechanical engineering | 125 | 0,1355 | 0,1401 |
| Industrial engineering | 81 | 0,1325 | 0,1514 |
| Nuclear engineering | 117 | 0,1606 | 0,1668 |
| Chemical engineering | 201 | 0,0996 | 0,1186 |
| Electronic enigineering | 210 | 0,1105 | 0,0904 |
| Telecommunication engineering | 135 | 0,1117 | 0,1203 |
| Bio-engineering | 110 | 0,1214 | 0,1332 |
| Informatics | 145 | 0,4052 | 0,4204 |
| Infrstructur engineering | 6 | na | na |
| **Area 13 Economics and Statistics** | **590** | **0,54** | **0,54** |
| Economics | 235 | 0,56 | 0,56 |
| Economic history | 37 | 0,32 | 0,29 |
| Management | 175 | 0,49 | 0,5 |
| Statistics | 143 | 0,55 | 0,55 |
| **All Areas** | **9199** | **0,32** | **0,38** |

**Source.** (ANVUR 2013). *Final Report*; Appendix B; Appendix A of each *Area Report*. All data

## 4. A meta-analysis of the experiment

Four meta-analyses were carried out, considering Cohen's kappas with either linear and VQR weights as well as comparing either areas or subareas. Indeed each (sub-)area can be considered as performing a parallel experiment on different subjects (Sun 2011; Spiegelhalter 2005).

The null hypothesis $H_0$ was formulated that the kappa of Area 13 is drawn from the same distribution as those from other areas. Recalling that Cohen's kappa is asymptotically normal and its sample variance is inversely proportional to the number $m_j$ of the assessed papers in the *j*-th area, under $H_0$ we have

$$k_j \sim N(\mu, \sigma^2/m_j) \ j = 1,2,\dots,n+1$$

where $\mu$ and $\sigma^2$ are unknown parameters characterizing the distribution; since there are 10 areas, we let $n=9$, while $\kappa_{10}$ denotes the kappa for Area 13. We assume that $k_j$ for $j = 1,2,\dots,9$ are observed. Then, the 95% prediction interval for $\kappa_{10}$ is derived and graphically represented as a funnel plot (technical details available as Online Resource). For both the linear- and VQR-weighted kappa, it is found that the point ($m_{10}$, $\kappa_{10}$) lies outside the funnel, meaning that $H_0$ is rejected with significance level 5% , see top panels of Fig. 1 for the funnel plots and Table 2 for the *p*-values.

The same procedure was applied to draw the funnel plots associated with sub-area kappas. Now the null hypothesis $H_0$ is that kappas of the 4 sub-areas within Area 13 are drawn from the same distribution as those of the 38 sub-areas belonging to the other 9 areas. Using the kappas of the 38 subareas, the 95% prediction interval is derived and graphically represented as a funnel plot. For both the linear- and VQR-weighted kappa, 3 out of 4 sub-area kappas of Area 13 lie outside the funnel, meaning that in these three cases $H_0$ is rejected with significance level 5%, see bottom panels of Figure 2 and also Table 3 for the *p*-values. Conversely, the kappa of the Economic history sub-area lies within the limits for both the linear- and VQR-weighted cases. Under $H_0$, the probability of 3 samples out of 4 exceeding the 95% prediction limits is less than $1.2 \times 10^{-4}$.

This further confirms the rejection of the hypothesis that the experiment within Area 13 had the same statistical properties as the experiments in the other 9 areas.

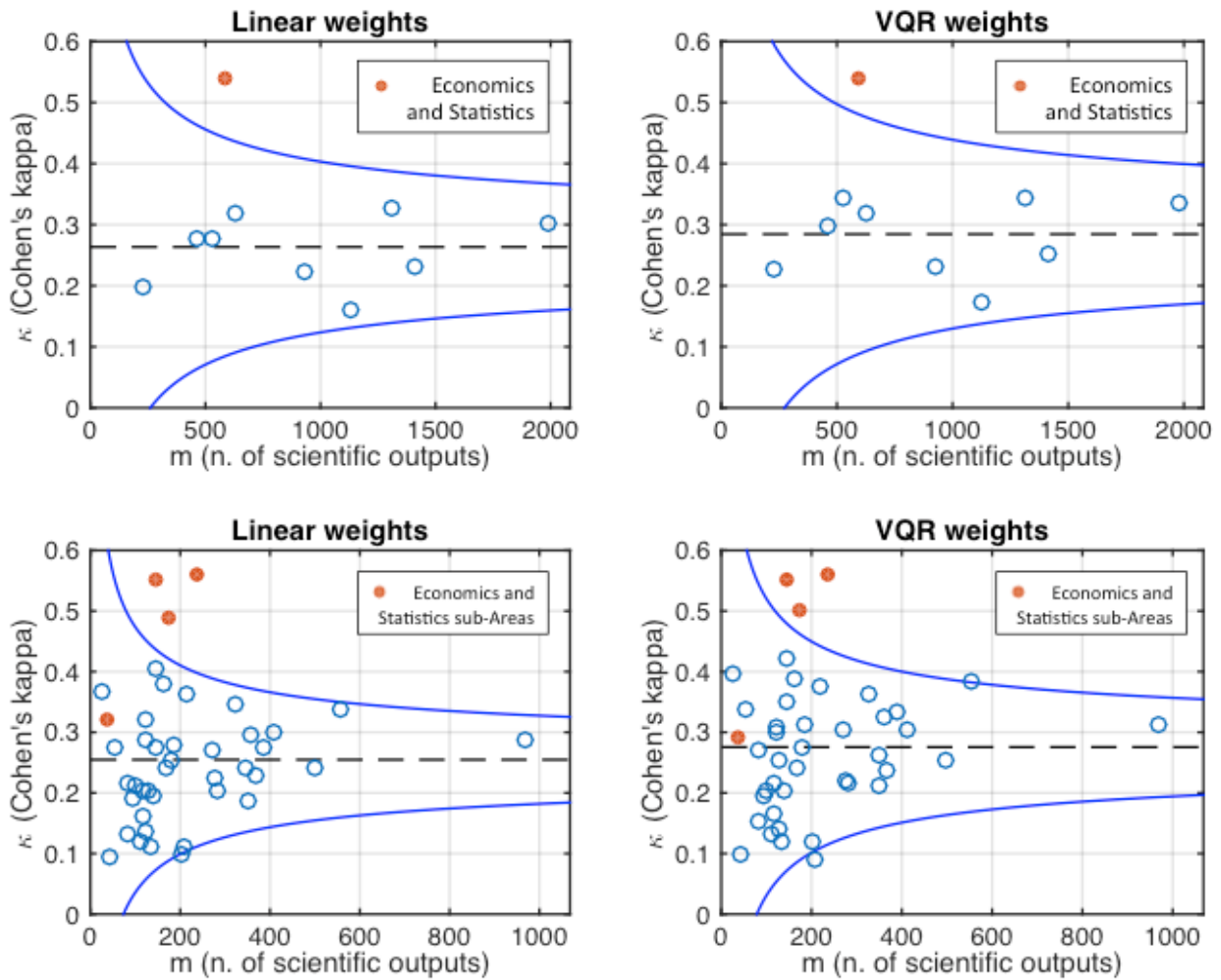

**Figure 2.** Funnel plots: a point with coordinates ($m$,κ) represents a (sub-)area having $m$ evaluated products and whose Cohen's kappa is κ. Cohen's kappas for Area 13 (full circles) are compared to the mean kappa (dashed) and 95% prediction limits (continuous), based on kappas collected in the other nine areas (open circles). Top: the kappas refer to the 10 areas. Bottom: the kappas refer to the sub-areas. Left: linearly-weighted kappas are considered. Right: VQR-weighted kappas are considered.

The computation of the prediction intervals performed for Area 13 was repeated also for all the other 9 areas, one at a time, by discarding kappas from the area to be tested and using the kappas of the other nine areas for estimating the 95% prediction limits. In all the nine cases, the tested areas had all their area and sub-area kappas lying within the prediction limits, except for Area 9, whose kappa for the Electrical and Electronic Engineering sub-area ($\kappa = 0.09$) fell below the lower prediction limit for the VQR-weighted case.

**Table 3.** *p*-values for Area 13 and its sub-areas.

|  | Linear weighted kappas | VQR weighted kappas |
|---|---|---|
| Area 13 Economics and statistics | 0.0036** | 0.0086** |
| *Sub-areas* | | |
| Economics | 0.0001*** | 0.0005*** |
| Economic history | 0.3571 | 0.4711 |
| Management | 0.0034** | 0.0096** |
| Statistics | 0.0012** | 0.0051** |

\* $p < 0.05$; \*\* $p < 0.01$; \*\*\* $p < 0.001$

As we can easily understand, Area 13 does not conform to the distribution of the other areas. Why does it happen? A first hypothesis is that in economics and statistics there are shared evaluations which do not exist in the other disciplinary areas. This hypothesis can be promptly rejected, given that a harsh discussion about research evaluation in economics have been developed worldwide also in reference to the British RAE (Lee 2007). A second hypothesis is that the experiment performed in Area 13 was substantially different from the ones of the other areas. In the following paragraph, we will argument in favour of this second hypothesis.

## 5. A self-fulfilling experiment

A careful scrutiny of ANVUR's *Final Report* and area reports permit to highlight at least five profiles by which Area 13 experiment is different from the ones performed in the other areas.

1. The first profile is about random sampling. The random sampling of articles in Area 13 was possibly affected by authors's requests of being evaluated through peer review.[22] No information about the extent of these insertions is currently available.

2. The second profile relates to the information available to reviewers. In Area 13, referees were in a position to be aware of their engagement in the experiment, while in all the other areas referees ignored to be engaged in the experiment.[23]

3. The third profile relates again to the information available to reviewers, but in this case about bibliometric evaluation. As we have seen, in the areas 1-9 bibliometric classification of articles was not easily available because it requested the complete knowledge of the suitable matrix; a referee could have foreseen the bibliometric classification of an article only with difficulty and with a quite high margin of error. In Area 13 the bibliometric classification of article was instead very easy because it was based on the journal ranking developed and published by the GEV. This ranking classified journal in the four VQR categories and each article was evaluated according to the classification of the journal in which it had been published. "The referees were provided with the panel journal classification list and the actual or imputed values of IF, IF5 [five years impact factor] and AIS [Article influence score]" (Bertocchi et al. 2015).[24] Therefore in Area 13 a referee requested to evaluate an article published in a journal ranked by the GEV, immediately knew the article's bibliometric evaluation: he just had to check the journal ranking. So in Area 13, not only referees knew that they were partecipating to the experiment, but they also knew the bibliometric evaluation with which their judgement was to be compared. These two pieces of information were not available to the referees ot the other areas.

---

[22] This point is clearly stated in the Appendix B of the Area Report: "The sample selection shall take account of any specific request for peer review reported via the CINECA electronic form for highly specialized and multidisciplinary products" (p.64). This information is not reported neither in (Bertocchi et al. 2015) nor in (Bertocchi et al. 2013e, 2013a, 2013b, 2013c, 2013d) or in (Ancaiani et al. 2015).

[23] As we have seen above, in the Areas from 1 to 9 the use of matrices may not give a definite result because a discordance between number of citations and impact factor. As a consequence, in these areas many journal articles were evaluated by IR. Referees of Areas 1-9 did not know if the article was sent them because of an uncertain bibliometric evaluation, or because the article was part of the random sample for the experiment. Indeed in Areas 1-9, to recognize that a research product belonged to the experiment sample, it was necessary to compare its citations with thresholds whose official values were never published by ANVUR. Instead, in Area 13 no journal articles were evaluated trough IR review except the ones of the experiment. So in Area 13 if a referee was requested to perform a peer review of a journal article, he or she may know immediately that he or she was engaged in the experiment.

4. The fourth and most critical profile relates to the procedure for reaching the final evaluation by using IR. As we have seen, the generic description of the experiment indicates that the scores separately assigned by the two referees were summarized according to an algorithm, in many cases a simple average of the two scores. In Area 13 the protocol was completely different. When the two referee's reports were comunicated to the two GEV members in charge of the considered article, they formed a "consensus group" which directly decided the final evaluation of the article, by considering the referee's reports as simple information for their final evaluation.[25] Under weak assumptions, it is possible to demonstrate that at least 326 articles out of 590 (55,3%) considered for the experiment were evaluated not by referees, but by the consensus groups, that is directly by the members of the GEV. In particular it is possible to estimate that GEV members evaluated directly at least 54.3% of articles classified as A (excellent); 58% of articles classified as B (good); 83.7% of the articles classified as C (acceptable), and 31.6% of the articles classified as D(limited). Details of these estimate are reported in the Appendix of this paper.[26]

---

[24] This information was not disclosed in the ANVUR's *Final Report*.

[25] If the opinion of the two referees coincided, the final evaluation was promptly defined. If the opinion of the two referees diverged, a complex process started. This process is stated in the appendix A of the Area 13 Report: "The opinion of the external referees was then summarized by the internal consensus group: in case of disagreement between [the two referee's reports], the [final score] is not simply the average of [the referee's scores], but also reflects the opinion of two (and occasionally three) members of the GEV13 (as described in detail in the documents devoted to the peer review process)" (ANVUR 2013). The work of the consensus groups is described as follows: "The Consensus Groups will give an overall evaluation of the research product by using the informed peer review method, by considering the evaluation of the two external referees, the available indicators for quality and relevance of the research product, and the Consensus Group competences." (ANVUR 2013). The consensus groups in some cases evaluated also the competences of the two referees, and gave "more importance to the most expert referee in the research field". (Area Report, p. 15 translation from Italian by the authors). It is worthwhile to note that in the overall Italian research assessment exercise the notion of "informed peer review" individuates at least two very different processes (ANVUR 2013). The first one refers to the evaluation made by external referees who knows metadata and bibliometric indicators about the refereed articles. The second one consists in the evaluation made by a consensus group, that is by two or more member of a GEV, who used not only the informed peer reviews produced by two referees, but also the bibliometric indicators available for the article, and a personal judgement about the competences of the referees (ANVUR 2013).

[26] It is worthwhile to recall that Area 13 panelists had decided the journal ranking used for bibliometric evaluation of articles submitted to the research assessment; they were the same panelists that decided the final score of the IR process. These modifications of the protocol, specific of the experiment performed in Area 13, have possibly introduced a substantial bias toward agreement between bibliometrics and IR. Indeed the relatively high kappas calculated for Area 13 can be interpreted as indicating a "fair to good agreement" between the evaluation based on the journal ranking developed by the panelists, and the IR performed by the panelists on the basis of the referee reports.

5. A fifth profile is connected to the two preceding ones and refers to the information available to GEV members in charge of choosing reviewers. GEV13 members knew that all journal articles which had to be submitted to peer review were used in the context of the experiment. They also knew with certainty for each article the merit class in which it was assigned by using bibliometric evaluation. In the other areas these two pieces of information were probably not so easily available to GEV members, because thay had many journal articles that had been submitted to peer review, possibly without a univocal bibliometric evaluation. It cannot be excluded that this information, available to GEV13 members, played a role in the choice of referees. In fact, a suitable choice of referees might have determined the fate of an article. Consider for example an article published in journal classified as Excellent, which applied standard techniques to a standard problem. If the GEV member desired to raise the probability that IR agreed with bibliometric evaluation, he or she had to choose a referee who notoriously appreciated the standard technique. A suitable choice of the referee by one of the panel member may also raise the probability that the article will be evaluated by a consensus group; in the preceding example, this end may be reached by choosing a referee notoriously critical toward the standard technique.

By considering these five profiles it appears that the experiment performed in Area 13 was radically different from those performed in the other areas. And that the higher livel of agreement between bibliometrics and IR in Area 13 may be the result of the modifications introduced by the Area 13 panel to the experiment protocol adopted in the other Areas.

A possible objection to this assertion relates to the subarea of "economic history",[27] where the adoption of Area 13 procedure did not result in relatively high kappa values. Why economic history, despite being subject to the Area 13 protocol, did not yield a relatively high value of kappa? Two conjectures may be made on this. The first is again about the protocol. Economic history is the only sub-area of the experiment for which a dedicated sub-area panel was not defined.[28] It may be conjectured that the Area panel might have encountered difficulties in enacting the appropriate choice of referees because there was not enough panelists who were expert in the economic history and history of economic thought field.[29] The second is about the idiosyncratic character of the economic history sub-area, which might have deserved a richer scrutiny, impossible with data publicly released by ANVUR. It is in fact one of the few sub-areas where bibliometric evaluations were lower than the ones obtained through IR. Moreover, as noted in the Area 13 Report, in the economic history sub-area "there was a higher disagreement between the two external referees, compared with other areas, and there were particularly strict or generous judgments. In some cases the discordant opinions between referees reflect the different point of views and approaches to the subject, as well as different quality judgments. In these cases the Consensus Groups were particularly useful and they managed to give importance, in different measures, to external referees' opinions and, consequently, to the most expert referee's point of view" (ANVUR 2013). These conjectures seem to suggest that the results of the experiment for economic history sub-area have to be considered with particular caution.

---

[27] Economic history is a short label for economic history and history of economic thought.
[28] As a rule, evaluations for each sub-area were conducted by a specific sub-area panel. Instead, journal articles classified as pertaining to "economic history" were evaluated by the panelist of the sub-area "economics". It is worthwhile to note that in the experiment the best agreement between IR and bibliometrics is reached exactly in a sub-area (economics) where a subset of observations with lower agreement, those of economic history, were treated separately from the others.
[29] In the Area 13 panel only one panelist is enrolled as full professor of economic history, and only one can be properly considered as an expert in the history of economic thought.

## 6. Concluding remarks

During the Italian research assessment exercise, the national agency ANVUR made an experiment in order to verify the agreement between results obtained through informed peer review and bibliometrics. A sample of articles submitted for the research assessment were evaluated by using both informed peer review and bibliometrics. The degree of agreement between IR and bibliometrics were then analyzed by using a statistical technique (the weighted Cohen's kappa). ANVUR official reports interpreted these results as indicating an overall "good" or "more than adequate" agreement between IR and bibliometrics. This result was widely disseminated as an alleged strong piece of evidence in the discussion about research evaluation techniques, by supporting the idea that peer review and bibliometrics are substitute because they give similar results. We have shown that when the available guidelines for interpreting kappa values are considered, for nearly all research areas and sub-areas considered , the degree of agreement (always in the range 0.09-0.42) has to be interpreted as "unacceptable", "poor" or in a few cases as, at most, "fair". The only notable exception is represented by Area 13 (economics and statistics) and its sub-areas, where the degree of agreement was in the range 0.29-0.56. The statistical meta-analysis performed, confirms that results for Area 13 are significantly different from those reached for the other areas.

A careful scrutiny of the experiment protocol adopted in Area 13 highlighted that the panel in charge of the evaluation introduced substantial modifications to the protocol adopted in all the other areas involved in the experiment. In Area 13, differently, from the other areas: (i) random sampling took into account authors' requests to be evaluated by peer review; (ii) the referees might have known that they were part of the experiment; (iii) the referees might have known the precise merit class in which each article was classified by using bibliometrics; (iv) the synthesis of the two referees's judgments was defined by a Consensus Group composed by (at least) two panel members, who considered the referees judgments as mere information for deciding the final score; (v) the panel members forming the Consensus Groups knew that their final judgment would be used for the experiment. As a consequence of some of these modifications, it can be estimated that at least 53% of the IR evaluations was not expressed by referees, but directly by the Area 13 panelists. For these reasons, results reached for Area 13 have to be considered as fatally flawed by virtue of the protocol modifications introduced by the area panel.

For the remaining Areas 1 to 9, it appears that the ANVUR experiment resulted in degrees of agreement between IR and bibliometrics that can be described as unacceptable, or poor to fair. This has relevant consequences on two different profiles.

The first one relates to soundness of results reached by the Italian research assessment exercise. As documented by ANVUR, articles evaluated through IR had lower evaluation than articles evaluated through bibliometrics (Cicero et al. 2013; ANVUR 2013). As a consequence, VQR final results reached for Areas, sub-areas, disciplinary sectors and institutions depend on the mix of instruments used to evaluate research outputs. *Ceteris paribus*, areas, disciplinary sectors or institution with a higher percentage of research outputs evaluated through IR, are more likely to have obtained lower average scores than areas, disciplinary sectors and institutions with lower percentage of IR. Negative correlations between overall results and the percentage of research outputs evaluated through the informed peer review were documented for two areas (Baccini 2014a; De Nicolao 2014). It is worthwhile to note that this negative correlation may originate by the mix of instruments used, but also by the intrinsic "lower quality" of research outputs which were evaluated by IR. Indeed, as we have seen, all research outputs different from articles in indexed journals were evaluated through peer review. The point to be underlined here is that it is impossible to disentangle these two phenomena, and as a consequence it is impossible to give a clear interpretation of the results of the Italian research assessment exercise. It is impossible to affirm, for example, in a comparison between two departments, if a department has a higher score because it had produced better research or because it had a lower percentage of research outputs evaluated by IR.

The second profile relates to the organization of large scale research assessment exercise. According to ANVUR, the point at stake in its experiment is the justification of the use of both bibliometrics and informed peer review in a same research assessment: "results of the analysis relative to the degree of concordance … may be considered to validate the general approach of combining peer review and bibliometric methods" (Ancaiani et al. 2015). The conclusion reached by ANVUR, according to which the "more than adequate" agreement between bibliometric and informed peer review "fully justifies" the use of both techniques of assessment, must be reversed. Indeed in the overall sample there is an unacceptable, a poor or at most, and in a few cases, a fair agreement, and this result all but justifies at all the joint use of both techniques. *A fortiori* the two techniques cannot be considered as "substitute", as sustained by (Bertocchi et al. 2013e, 2013a, 2013b, 2013c, 2013d). Not even this policy conclusion drawn for the experiment performed in Area 13 appears to be well founded: "the agencies that run these evaluations could feel confident about using bibliometric evaluations and interpret the results as highly correlated with what they would obtain if they performed informed peer review" (Bertocchi et al. 2015), since the good agreement in this area was reached by a manipulated experiment.

Indeed, the Italian research assessment exercise and the experiment conducted by the ANVUR seem to support that informed peer review and bibliometric analysis do not produce similar results. This suggest that the agencies that run research assessments should feel confident that results that will be reached by using a technique will differ from those reached using the other. A result coherent with previous literature and with evidence provided for the REF by the correlation analysis performed in the context of the "Independent review of the role of metrics in research assessment and management"

**Appendix**

In Area 13, 590 journal articles were selected for the experiment. Each article was assigned to two GEV members responsible of the IR process. Each of the two GEV members chose an external referee. After having received the referees scores, the two GEV members met up in a Consensus group in charge of defining the final IR evaluation of the paper. If we suppose (Hypothesis 1) that none of the articles was evaluated at a first sight as D (limited) by both GEV members –in this case an article was not submitted to the IR process-, then the 590 articles products can be treated as they were evaluated by two referees. Let's suppose now (Hypothesis 2) that the Consensus Group formed by the two GEV members has never modified a concordant judgement, i.e. a judgment for which both referees were in agreement. We know (ANVUR 2013) that the referees gave concordant judgments for 264 articles. This means that the evaluation of, at least, 326 articles, that is 55.3% of the total, was decided by Consensus Groups composed by GEV members. Table A1 summarizes specific estimates for the merit classes.

Table A1. Distribution of articles per merit classes and estimate of the number of articles evaluated directly by the Area 13 panelists.

|  | Merit classes | | | | |
| --- | --- | --- | --- | --- | --- |
|  | A | B | C | D | TOTAL |
| (1) # papers Bibliometric evaluation | 198 | 102 | 103 | 187 | 590 |
| (2) # papers Informed peer review | 116 | 174 | 129 | 171 | 590 |
| (3) # papers with concordant peer reviews | 53 | 73 | 21 | 117 | 264 |
| (4) Concordant biblio and peer ev. | 98 | 56 | 39 | 118 | 311 |
| (5=2-3) # papers evaluated by GEV members | 63 | 101 | 108 | 54 | 326 |
| (6=5/2) % of papers evaluated by GEV members | 54.3% | 58.0% | 83.7% | 31.6% | 55.3% |
| (7=4-3) # papers evaluated by GEV concordant with biblio | 45 | -17 | 18 | 1 | 64 |

**Source:** own elaboration on (ANVUR 2013). A slightly different version was published in (Baccini 2014b).

In the columns we can read the merit classes. Rows 1 and 2 contains respectively the distribution of articles per merit classes as judged by using bibliometrics and IR. Row 3 shows the number of articles for which the two referees expressed a concordant evaluation. In the row 4 the numbers of articles for which the IR was in agreement with the bibliometric evaluation are reported. Row 5 contains the estimate, under Hypothesis 1 and 2, of the minimum number of articles whose final IR evaluation was decided by the GEV Consensus Groups. We can see that 54.3% of articles classified as A through IR were evaluated directly by the GEV Consensus Groups; this percentage increases to 58% for articles classified as B; and at least 83.7% of articles classified as C were decided by Consensus Groups. The percentage decreases to 31.6% for products evaluated as D.

It is worthwhile to note that the hypotesis 1 and 2 tend to lower the estimate of the number of articles directly evaluated by the Consensus groups. In particular, for this estimate we are assuming that in Area 13 GEV members never agreed to directly evaluate as D an article, in which case it was scored D without being sent out for peer review. Consider that in the other areas, the percent of articles that received a concordant D score by two referees is 21.1% (705/3441). In Area 13 this percentage is more than doubled: 44.3% (117/264). It is impossible to establish, by using publicly available data, how much this result is due to a higher level of agreement between the two referees or to an initial agreement of the two GEV members, both evaluating an article as D. On this basis, we can presumptively affirm that the number 54 is just an underestimate of the number of articles evaluated as D by the Consensus Group.

A third hypothesis is also formulated: each time peers expressed a concordant evaluation, this coincided with the bibliometric evaluation. For example: every time two referees agreed to judge an article as A, the work resulted classified as A also by using bibliometrics. Under this hypothesis it is possible to estimate (row 7 of the table) the minimum number of articles for which the Consensus group decided an evaluation coincident with the one reached by bibliometrics. At least 64 articles out of the 311 (21%) for which IR and bibliometrics evaluation coincided were directly evaluated by the Consensus Group. This value is strongly underestimated. Indeed, for articles evaluated as B, peers were in agreement for 73 articles; but bibliometric and peer review coincided for just 56 articles.